\documentclass[aps,pra,superscriptaddress,twocolumn]{revtex4}
\usepackage{amsmath}
\usepackage{calligra}
\usepackage{amsfonts}
\usepackage{graphicx}
\usepackage{dcolumn}
\usepackage{stackrel,amssymb}
\usepackage{color}
\usepackage{ulem}

\usepackage{bm}

\begin{document}

\title{Completely subradiant multi-atom architectures through 2D photonic crystals}
\author{F. Galve}
\email{fernando@ifisc.uib-csic.es}
\affiliation{IFISC (UIB-CSIC), Instituto de F\'isica Interdisciplinar y Sistemas 
Complejos, Palma de Mallorca, Spain}
\author{R. Zambrini}
\affiliation{IFISC (UIB-CSIC), Instituto de F\'isica Interdisciplinar y Sistemas 
Complejos, Palma de Mallorca, Spain}

\begin{abstract}
Inspired by recent advances in the manipulation of atoms trapped near 1D waveguides and proposals to use surface acoustic waves
on piezoelectric substrates for the same purpose, we show the potential of two-dimensional platforms. 
We exploit the directional emission of atoms near photonic crystal slabs with square symmetry to build perfect subradiant 
states of 2 distant atoms, possible in 2D only for finite lattices with reflecting boundaries. We also show how to design massively parallel
1D arrays of atoms above a single crystal, useful for multi-port output of nonclassical light, by exploiting destructive interference 
of guided resonance modes due to finite size effects. Directionality of the emission is shown to be present whenever a linear
iso-frequency manifold is present in the dispersion relation of the crystal. Multi-atom radiance properties can be obtained 
from a simple cross-talk coefficient of a master equation, which we compare with exact atom-crystal dynamics, showing its predictive power.
\end{abstract}

\maketitle

Engineering the interaction between discrete and continuous variable degrees of freedom has proven a key ingredient for many quantum information
processing platforms. Prominent examples are e.g. 2D Coulomb crystals of trapped ions, where
vibrational quanta of the ions' motion are used as channel to effectively induce spin-spin interactions with tunable 
distance dependence \cite{thompson,Bollinger1,Bollinger2}. Tuning long range couplings in 1D ion chains, has allowed for propagation of correlations
in the regime where the effective-light-cone picture does not apply \cite{monroe2014,blat2014}.
Recently, excellent control of interactions between atoms and light in 1D waveguides has been shown \cite{painter2014}, leading to observation
of many atoms superradiance \cite{painter2015}. This has spurred theoretical proposals to use the engineered properties of light to mediate
strong long-range atom-atom interactions both in 1D \cite{chang2015} and 2D photonic crystal (PC) lattices \cite{tudela2015,tudela2016}. 
Another promising platform is based on surface acoustic waves (SAW) on piezoelectric 
materials \cite{giedke2015,giedke2017}, acting as mediators between e.g. quantum dots, trapped ions, nitrogen-vacancy centers, or superconducting qubits.
While coherent interactions are useful for quantum simulations, radiative interactions not only expand the toolbox of quantum optics, but have 
important practical implications for quantum networking \cite{qinternet1,qinternet2} or quantum memories \cite{qmem1,qmem2,qmem3}.

In this Letter we focus on emitters coupled to 2D periodic structures, which can be described as tight-binding models for photons or phonons, because they allow 
richer features than 1D lattices. The physics of this problem is generic enough that it can be applied to e.g. 2D Coulomb crystal of trapped 
ions \cite{thompson,Bollinger1,Bollinger2}, PCs with nearby trapped atoms \cite{painter2014,painter2015,chang2015,tudela2015}, 3D printed 
photonic circuits \cite{Sciarrino1,Sciarrino2}, trapped-ions near piezoelectric substrates \cite{giedke2015, giedke2017}, cold atoms in 
optical lattices \cite{ines}, or even circuit QED architectures \cite{cQED0,cQED1,cQED2,cQED3}. We focus on directional emission in 2D periodic 
lattices \cite{Galve2016,2Dphotonic}, explaining its origin and generality. This is exploited, together with reflecting boundaries \cite{Bragg}, 
to show a great variety of dark-state multi-atoms configurations. These architectures are in principle amenable to be used for the creation of steady state many-atom entanglement
\cite{tudelaPorras}, to create multi-photon states in a controlled way \cite{Tudela3} and expands to 2D platforms the toolbox of atom-light quantum optics. In particular we show how to 
build, %for a lattice with $N\times N$ sites, dark states of $2N+3$ atoms in a configuration where atoms are radiatively coupled in sets of four atoms, while 
in a rectangular configuration, a dark state of $\sim N_xN_y/4$ atoms formed by radiatively independent `lines' of $\sim N_x/2$ atoms each. The latter is equivalent 
to a multi-1D setup, useful for parallel multi-output generation of N-photon states \cite{TudelaMultiphoton}.
\begin{figure}[h]
\includegraphics[width=\columnwidth]{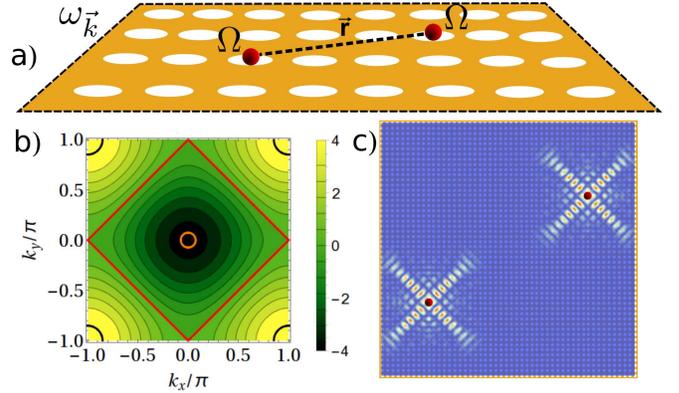}
\caption{a) Sketch of a periodic structured substrate, a 2D lattice (e.g. a photonic crystal), with reflecting boundaries.
Two, or more, emitters with energy splitting $\Omega$ couple locally to the structure at a relative distance $\vec{r}$. 
b) The lattice's dispersion relation $\omega_{\vec{k}}$ (in units of $J$) regulates which manifold of excitations $\vec{k}_\Omega$ is resonant
with the emitters frequency ($\Omega$). For $\Omega$ close to the band's frequency edges, black and orange lines, isotropic emission ensues. We will
work with $\Omega=0$, i.e. the middle of the band (red line), where directional emission is known to occur \cite{Galve2016,2Dphotonic}.
c) Example of directional emission pattern for $\Omega=0$, and initial state of two emitters $|\psi\rangle=(|01\rangle+|10\rangle)/\sqrt{2}$. We plot
the population $|c_{\vec{r}}|^2$ of each lattice's site at early times.} 
\label{figDIBUJO}
\end{figure}\\
\indent {\it Finite lattice.--} We study the interaction of a two-unit, bosonic or atomic, quantum system coupled to a periodically structured
bosonic substrate such as that of Fig.~\ref{figDIBUJO}a (a PC lattice is represented).
For a square-symmetry tight-binding lattice the dispersion relation reads $\omega_{\vec{k}}=-2J(\cos k_x +\cos k_y)$, Fig.~\ref{figDIBUJO}b.
Two identical and independent emitters can be either bosons, $H_S=\Omega(a_1^\dagger a_1+a_2^\dagger a_2)$, or  two-level atomic systems (TLS), $H_S=\Omega(\sigma_1^+ \sigma_1^-+\sigma_2^+ \sigma_2^-)$.
We consider local coupling to the structure $H_{\text{int}}=\lambda(\sigma_1^- A(\vec{r}_1)^\dagger+\sigma_2^- A(\vec{r}_2)^\dagger +h.c.)$ 
[from now on, the bosonic case follows directly by substitution of $\sigma_j^-\to a_j$ and $\sigma_j^+\to a_j^\dagger$], with $A(\vec{r})=\sum_{\vec{k}}f_{\vec{r},\vec{k}} A_{\vec{k}}$ and 
$H_{\text{latt.}}=\sum_{\vec{k}}\omega_{\vec{k}} A_{\vec{k}}^\dagger A_{\vec{k}}$, where $f_{\vec{r},\vec{k}}$ is the spatial profile of guided mode $\vec{k}$ in the lattice (see below).
The  annihilation operators $A(\vec{r})$ for the light field (atoms interacting with PC) or surface displacement field (SAW case),
are evaluated at the positions of the emitters $\vec{r}_1$ and $\vec{r}_2$.
In the case of atoms coupled radiatively to a photonic crystal we will study the 1-photon case. This case already shows many interesting features and a rich variety of dark state configurations. 
In the single-excitation sector, the exact dynamics can be computed numerically (see \cite{Tudela1} for analytic results in the infinite lattice case) even for strong atoms-lattice coupling, by the Schr\"odinger equation
$i\partial_t|\psi_t\rangle=H|\psi_t\rangle$ (we set $\hbar=1$) and noting that the total state of e.g. two emitters is $|\psi_t\rangle=c_1(t)|10,\vec{0}\rangle+c_2(t)|01,\vec{0}\rangle +\sum_{\vec{k}}c_{\vec{k}}(t)|00,\vec{k}\rangle$
(with the single excitation being in emitter 1, 2, or in mode $\vec{k}$ of the lattice, respectively). The real-space amplitude in the lattice is $c_{\vec{r}}(t)=\sum_{\vec{k}}f_{\vec{r},\vec{k}}c_{\vec{k}}(t)$, Fig.~\ref{figDIBUJO}c,.
We show that the full dynamics is easily 
understood in terms of a master equation, valid only in the weak atoms-lattice coupling limit. 
After a standard Born-Markov treatment \cite{breuer}, the dynamics with a lattice at T=0, is given by
\begin{equation}
 \dot{\rho}_S=-i[\tilde{H}_S,\rho_S ]+\sum_{j,l=1}^2 \Gamma_{j l}(t) ( \sigma_j^- \rho_S \sigma_l^+ - \frac{1}{2} \{\sigma_l^+ \sigma_j^-,\rho_S\})
\end{equation}
with $\tilde{H}_S=H_S+H_{LS}$ the Lamb-Shift corrected Hamiltonian, $H_{LS}=\Omega_{LS}(\sigma_1^+ \sigma_1^-+\sigma_2^+
\sigma_2^-)+\lambda_{LS}(\sigma_1^-\sigma_2^++h.c)$. We define the cross-talk $\Gamma_c\hat{=}\Gamma_{12}=\Gamma_{21}$ and the 
individual decay rates $\Gamma_{11}$, $\Gamma_{22}$. After a very short time compared to the inverse atoms-lattice coupling, all coefficients in the master equation reach a 
constant value \cite{Galve2016}, and the cross talk becomes $\Gamma_{j l}=\lambda^2\sum_{\vec{k}_\Omega}f_{\vec{r}_j,\vec{k}_\Omega}f_{\vec{r}_l,\vec{k}_\Omega}^*$, with $\vec{k}_\Omega$ 
denoting the lattice's pseudomomentum manifold resonant with $\Omega$ \cite{GalveArxiv}.
Modelling the PC in Fig.~\ref{figDIBUJO}a by a tight-binding approximation with reflecting boundary, we take $f_{\vec{r},\vec{k}}=2/(N+1)\sin[k_x x]\sin[k_y y]$, with $k_\alpha=\pi l_\alpha/(N+1)$ and $l_\alpha\in[1,N]$.
Such reflecting boundary conditions have been experimentally demonstrated in \cite{Bragg}, by using Bragg mirrors.
Note that the cross-talk $\Gamma_{1 2}$ is the 1-photon propagator between positions $\vec{r}_1$ and $\vec{r}_2$ and thus also gives a good idea of what the emission pattern of atom $i$ is at 
point $\vec{r}_j$ (and vice versa). Its time dependence accounts for the build-up of a communication bridge between quantum emitters \cite{Benedetti, Galve2016}. 
In the case of a finite-sized lattice, individual decay rates depend on the position of the emitters but after a negligible transient stabilize around $\Gamma_{11}\simeq\Gamma_{22}=:\Gamma_0$. 
The dynamics is then diagonal in the operator basis $\sigma_\pm^-=(\sigma_1^-\pm\sigma_2^-)/\sqrt{2}$, 
$\dot{\rho}_S=-i[\tilde{H}_S,\rho_S ]+\sum_{j=\pm} \Gamma_{j}( \sigma_j^- \rho_S \sigma_j^+ - \{\sigma_j^+ \sigma_j^-,\rho_S\}/2)$
with new decay rates $\Gamma_\pm=\Gamma_0\pm\Gamma_c$ and 
$\tilde{H}_S=\sum_{\pm}(\Omega+\Omega_{LS}\pm\lambda_{LS})\sigma_\pm^+ \sigma_\pm^-$. Thus $|\Psi_\pm\rangle=(|01\rangle\pm|10\rangle)/\sqrt{2}$ is a dark state of the dissipator if $\Gamma_c/\Gamma_0=\mp 1$.
For continuous variables in the bosonic case $\Gamma_c=\pm\Gamma_0$ is known to lead to preservation of entanglement asymptotically \cite{pazroncaglia,Galve2016}. 
The closest Gaussian state analogue to the atomic Bell states consist in having one quadrature $x_\pm$ in vacuum and the other one $x_\mp$ squeezed.
If the squeezed quadrature has zero decay rate it will lead to preserved entanglement.

\begin{figure}[t]
\includegraphics[width=\columnwidth]{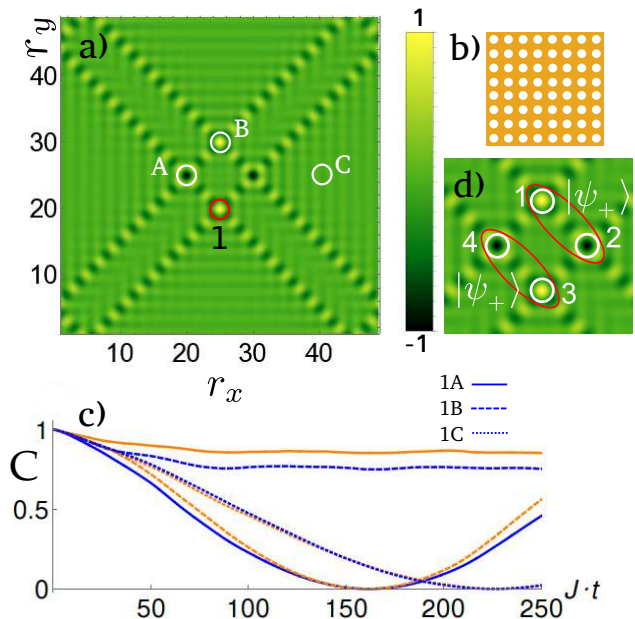}
\caption{a) Cross-talk function $\Gamma_{12}$ between emitter 1 (red sphere) at $\vec{r}_1=(x_1,y_1)=(25,20)$ and other emitter at $(r_x,r_y)$.
Dark(light)-color value at point A(B) is $\Gamma_{12}/\Gamma_{11}=+(-)1$. Diagonal directions have $\Gamma_{12}/\Gamma_{11}=\pm 1/2$ and $\sim 0$ elsewhere (e.g. in point C).
Atom positions are $\vec{r}_A=(20,25)$, $\vec{r}_B=(25,30)$, $\vec{r}_C=(40,25)$. b) Sketch of the photonic crystal orientation.
c) Exact dynamics of two atoms, initialized in Bell states $|\psi_+\rangle$ (orange) and $|\psi_-\rangle$ (blue). We plot concurrence \cite{conc} (a measure of two-qubit entanglement)
in time for the second emitter being at points A continuous lines, B dashed lines and C dotted lines. In cases A and B states $|\psi_+\rangle$ and 
$|\psi_-\rangle$ survive for long times, with an obvious retardation effect for B: waves have to propagate longer to bridge between emitters, and 
during this time ($J\cdot t\sim 40$) they decay independently; after that $\Gamma_{12}$ reaches a value equal to the decay rates. For the case C 
$\Gamma_{12}\sim 0$ and both Bell states decay equally fast. The atoms-lattice coupling is $\lambda=0.05J$.
d) Zoom in, now with four atoms. A simple dark state can be created by superposing two-atom dark states $|\psi_+\rangle$ of atoms 1 and 2, and 3 and 4. More complicated dark states of four atoms are also possible (see main text).} 
\label{FIG1}
\end{figure}

{\it Radiative directional coupling.--} When emitters are resonant with the middle of the band $\Omega=0$ they display purely directional emission along the diagonals of the lattice
(see Figs.~\ref{figDIBUJO}c, \ref{FIG1}a) \cite{2Dphotonic,Galve2016,Tudela1}, a scenario in stark contrast with the widely considered isotropic case.
This pseudomomentum manifold $k_y=\pm(\pi-|k_x|)$ contains four van Hove singularities \cite{vanHove53} in the infinite lattice case and leads
to divergent decay rates within perturbative treatments \cite{Tudela1}. In the finite lattice case there is no such problem. For a square-symmetry 
lattice the origin of directional emission is the presence of a frequency-degenerate iso-energy manifold of momenta which presents a {\it straight line} shape 
in momentum space, Fig.~\ref{figDIBUJO}b. Using the coordinates $k_\pm=(k_x\pm k_y)$, we can fulfill $\omega_{\vec{k}}=\Omega=0$ for $k_+=\pi$ and any $k_-$. 
This implies that along direction $x_+$ the cross-talk is a periodic (non-decaying) function
, whereas along $x_-$ we have an incoherent sum of many periodicities related to each $k_-$, and thus a decaying function. This is easiest to see in an infinite crystal, 
where radiation through plane waves would result into the sum $e^ {i \pi x_+}\int dk_- g(k_-)e^ {i k_- x_-}$, with $g(k_-)$ the density of states with momentum $k_-$, i.e.
a plane (non-decaying) wave along $x_+$ and an incoherent sum of plane waves in direction $x_-$, yielding a decaying function. Thus, rather generically, directional emission exists 
whenever there is a straight iso-frequency manifold. This happens for square and triangular lattices (also their higher-dimensional analogues) as well as for graphene \cite{graphene}.
Notably, introducing a small hopping term beyond first neighbors along diagonals of the lattice leads to $\omega_{\vec{k}}=-2J(\cos k_x +\cos k_y)-4\tilde{J}\cos k_x \cos k_y$. 
The last term introduces curvature in the formerly linear shape of the iso-energy manifold, leading to shorter-ranged radiation pattern. Hence, even for a periodic lattice directionality
is not immediate.

{\it Complete subradiance.--} Dark states (perfect subradiance) can arise among distant atoms in 1D where $\Gamma_c^ {(1D)}/\Gamma_0=\cos(k_\Omega x)$ \cite{painter2015,Galve2016} due to wave confinement.
In 2D isotropic media $\Gamma_c^ {(2D)}$ decay with distance. Furthermore, it was noted in a recent paper \cite{Tudela1} that even in structured infinite 2D media, 
no subradiant configuration of two atoms exists. Even in the case where directional emission is exploited, destructive interference is only possible along the `line-of-sight' between atoms, whereas the 
orthogonal direction can emit freely (see Fig.~\ref{figDIBUJO}c). Thus the best situation one can obtain is $\Gamma_c^ {(2D)}/\Gamma_0=1/2$ for two atoms, or perfect subradiance for four atoms\cite{Tudela1} (which now can cancel
both propagation directions). Reflecting boundaries \cite{Bragg} in the configuration of Fig.~\ref{FIG1}b, however, allow for closing of both emission lines, Fig.~\ref{FIG1}a, and thus
total cancellation ($\left|\Gamma_c/\Gamma_0\right|= 1$) at four points, allowing for complete subradiance for atom pairs. In Fig.~\ref{FIG1}a,c we show the exact dynamics of atoms, in agreement with the 
prediction from the cross-talk function. Apart from a retardation effect $J\cdot t\lesssim 40$, due to the buildup of $\Gamma_c$ which requires wave propagation between both emitters,
dark states can be created at arbitrary distances.
 
\begin{figure}[h]
\includegraphics[width=\columnwidth]{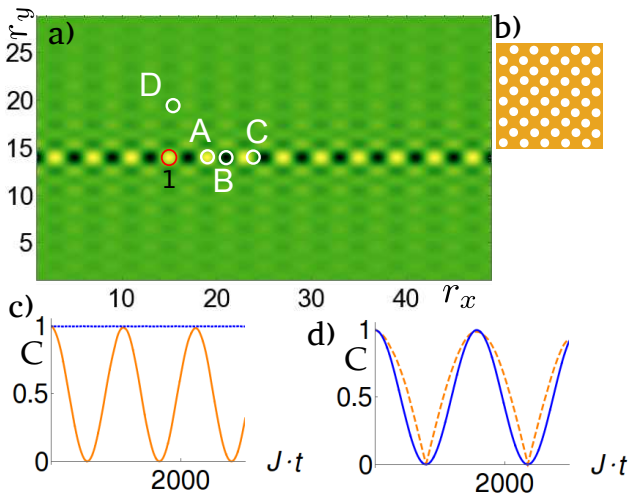}
\caption{a) Cross-talk function $\Gamma_{12}$ between emitter 1 at $\vec{r}_1=(15,14)$ and other emitter at $(r_x,r_y)$, with a photonic crystal of dimensions $N_x=49$ and $N_y=29$.
b) Sketch of photonic crystal symmetry (tilted). Color code as in Fig.~\ref{FIG1} . Lower row: exact dynamics of concurrence of two atoms with 
the second atom at $\vec{r}_2$: A $(19,14)$, B $(21,14)$, C $(24,14)$ and D $(15,20)$.  c) Initial state in A is $|\psi_+\rangle$ (orange, continuous) and $|\psi_-\rangle$ (blue, dashed). Curves for B are $|\psi_-\rangle$ (orange),
$|\psi_+\rangle$ (orange). d) Curves for C (orange, continuous) and D (blue, dashed). Both initial states $|\psi_+\rangle$ and $|\psi_-\rangle$ radiate equally. The atoms-lattice coupling is $\lambda=0.01J$.} 
\label{FIG2}
\end{figure}

{\it Tilted configuration.--} The remarkable interplay of geometry and boundaries can be appreciated by rotating the square-symmetric lattice by 45$^{\circ}$, Fig.~\ref{FIG2}b. The dispersion 
relation now reads $\omega_{\vec{k}}=-2J\cos k_x \cos k_y$ and directional emission occurs along horizontal and vertical lines. There is however some 
peculiarities associated to this configuration: first, from the argument given before, note that directional emission in e.g. the horizontal axis $x$ requires that 
$\omega_{\vec{k}}$ is constant for a given $k_x$ at all $k_y$, and remember that $k_x$ is discrete. The only way to have constant $\omega_{\vec{k}}$
for some value of $k_x$ irrespective of $k_y$ is to make $\omega_{\vec{k}}$ zero, which happens for $k_x=\pi/2$, i.e. $l_x=(N_x+1)/2$. This can only be fulfilled 
if $N_x$ is odd, otherwise $l_x\notin\mathbb{N}$. Note that in the non-tilted configuration we do not have this problem \footnote{There the dispersion is $\omega_{\vec{k}}=-2J(\cos k_x+ \cos k_y)=-4J\cos(k_+/2)\cos(k_-/2)$
so we need to make e.g. $k_+=\pi$. Since $k_+=\pi(l_x+l_y)/(N+1)$, one can always fulfill $l_x+l_y=N+1$ both for $N$ odd and even.}.
So we can only have a 'good' directional emission pattern for odd number of sites. We will in the following consider a rectangle with $N_x\neq N_y$, but both 
of them odd (a rectangular configuration in the non-tilted case would lead to multiple bounces, beyond the scope of this work).
The second peculiarity is that directional emission can only occur if one atom is placed in an odd position with respect to 
the wavevector which causes directionality: take e.g. directional emission along $x$, which needs $k_x=\pi/2$ ($l_x=(N_x+1)/2$), then$f_{\vec{n},\vec{k}}\propto\sin(\pi x/2)\sin[\pi l_y y/(N+1)]$ with $x\in\mathbb{N}$.
Thus whenever the $x$ of one atom is at even sites, it has null overlap with the modes possessing directional emission, which is thus suppressed in that direction. 
This can be used to devise linear emission patterns, as in Fig.~\ref{FIG2}, where two atoms see a cross-talk $\Gamma_c/\Gamma_0=\sin(\pi x_1/2)\sin(\pi x_2/2)$, i.e. 
alternating sign every 2 sites (and null cross-talk if any one of them is at an even site).
From the behavior of concurrence \cite{conc} in Fig.~\ref{FIG2} we see we can place atoms periodically along a line to have a complete dark state. Two-atom states that are not dark see a revival of their concurrence, 
due to multiple roundtrips of light, which ultimately will lead to total loss due to the finite quality factors of Bragg mirrors at the boundaries. In stark contrast dark states are much more robust 
against losses \cite{Tudela2} because the crystal remains unpopulated (although due to retardation effects the population is never exactly zero).

{\it Multi-unit architectures.--} We show next two possible configurations to demonstrate the richness of subradiant configurations that are possible in 2D structured media.
A simple way to design dark states of many atoms is by diagonalization of the dissipation matrix $\Gamma_{ij}$, similarly to the case of only two emitters described before. The Lindblad equation is then
diagonal and one can build linear combinations of its dark states to build others. Dark states of the dynamics, written as $|\psi\rangle=\sum_j c_j\sigma_j^+|00..0\rangle$ fulfill ${\mathbf \Gamma} \vec{c}=0$.  
The first configuration is represented in Fig.~\ref{FIG1}d: the four atoms have $\Gamma_c/\Gamma_0=+(-)1$ for equal(different)-colors, so the dissipation matrix is 
${\mathbf \Gamma}=\Gamma_0 \{\{1, -1, 1, -1\}, \{-1, 1, -1, 1\}, \{1, -1, 1, -1\}$ $,\{-1, 1, -1, 1\}\}$. One can proceed to diagonalization of this matrix, and take as dark states any combination of its eigenvectors 
with null eigenvalues; an example is the four particle dark state is $\{1,3,1,-1\}/(2\sqrt{3})$, with vector notation in the basis $|1000\rangle$, $|0100\rangle$, $|0010\rangle$, $|0001\rangle$ (with $|1000\rangle=\sigma_1^+|0000\rangle$).
Another, more intuitive, way is to take pairs of Bell states known to be dark in the two-atom case and superpose them, as in Fig.~\ref{FIG1}d:
$|\psi_{\text dark}\rangle=(|\psi_+\rangle_{12}+|\psi_+\rangle_{34})/\sqrt{2}$. This gives a correct dark state of four particles. We have checked with the exact dynamics that indeed these states are completely dark, 
up to retardation effects \footnote{It is peculiar that the  latter dark state looses less population by retardation effects than the superposition of Bell states.}.
\begin{figure}[h]
\includegraphics[width=0.9\columnwidth]{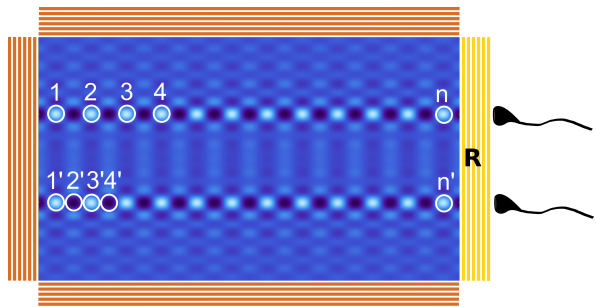}%{4atoms_caseAB.eps}
\caption{Joint representation (for illustrative purposes) of the cross-talk functions: {\it  upper line}, $\Gamma_{1,j}$ between emitter 1 at $\vec{r}_1$ and other emitter at $\vec{r}_j$;
{\it lower line}, $\Gamma_{1',j}$ between emitter 1' at $\vec{r}_{1'}$ and other emitter at $\vec{r}_j$.
Upper and lower emitters have zero cross-talk and can be used as independent 1D architectures. A lower Bragg reflectivity $R\lesssim1$ can be used to output light to optical fibers (right).
The upper row yields a Dicke model, whereas the lower row results in an alternate-sign Dicke model. } 
\label{FIG3}
\end{figure}

To fill the plane with as many atoms as possible and still keep a dark state, we can fill the horizontal and vertical in the form of a cross, i.e. $((N+1)/2,i)$ and $(i,(N+1)/2)$, $i=1...N$, except for the central position.
Every new square of four atoms we add, fulfills independently the conditions that allow superposing two Bell states, as before. We can further add an atom at the central position and
four more atoms at the vertices of the square. This configuration, with a total of $2(N-1)$ (the cross) plus $5$, i.e. $2N+3$ atoms, fills the lattice with cross-talk `paths' 
of value $1/2$, so one could not add any more atoms and keep a totally dark state. 

{\it Multi-1D waveguide array.--} We can use the tilted configuration, Fig.~\ref{FIG2}b, to produce dark states of linear arrays of atoms, where each array is radiatively {\it independent} of the other.
A four atom dark state (see Fig.~\ref{FIG3}) can be created with atoms 1, 2, 1', 3' by the superposition $(|\psi_-\rangle_{12}+|\psi_-\rangle_{1'3'})/\sqrt{2}$. 
If we modify now atom 3' to have $(|\psi_-\rangle_{12}+|\psi_+\rangle_{1'3'})/\sqrt{2}$, it would become superradiant only along the lower line. If we place atoms only in the 
upper line of Fig.~\ref{FIG3} at the light-color positions, we would have a Dicke model \cite{Dicke} $\dot{\rho}=\Gamma_0 (S^-\rho S^+ -\{S^+S^-,\rho\}/2)$, 
with $S^-=\sum_{i=1}^n\sigma_i^-$. If instead we fill the line as the lower one of Fig.~\ref{FIG3} we would have the jump operator $S^-=\sum_{\text{even}}\sigma_i^--\sum_{\text{odd}}\sigma_i^-$.
Implementation of a Dicke model allows e.g. for generation of many-atom entangled steady states \cite{tudelaPorras} and of many-photon single-mode states \cite{TudelaMultiphoton}.
Note that $(N_y-1)/2$ lines can be built in, with $\sim N_x/2$ atoms in each, so a dark state with a total of $\sim N_xN_y/4$ atoms can be created. 
Further, Fig.~\ref{FIG3} shows the sketch to use this idea as a parallel multi-1D setup, where each row of atoms behaves as a 1D array.
Protocols designed for 1D arrays of atoms can be implemented in parallel $\sim N_y/2$ times, using guided resonant modes \cite{JoanoGUIDED,Bragg}, with light being collected at a multi-port output,
equivalent to the leaky part of a cavity. We finally remark that the Lamb-Shift interaction $\lambda_{LS}$ is zero for all the cases studied here \footnote{It is not only much smaller than any other 
radiative coefficient, but has the same spatial shape as the cross-talk. That is, the Lamb-shift matrix and dissipation matrices are equal
up to a factor, and because dark states satisfy $\Gamma\cdot\vec{c}=0$, they are also left {\it invariant} by the Lamb-shift.}.

{\it Discussion and outlook.--} We have exploited the radiative directional interaction of emitters induced by a 2D periodic structure with reflecting boundaries. 
We show that the directionality is caused by frequency-degenerate linear-shaped manifolds in the substrate, which are expected to 
be of generic importance in 2D dispersion relations, although we present a specific mechanism that can break the linear shape in our particular case.
A master equation analysis is shown to properly describe the relevant emission phenomena, as compared to an exact dynamics treatment.
In stark contrast with an open-boundary 2D lattice, we find completely subradiant configurations of two atoms. We have proposed several multi-atom subradiant arrangements. In
particular, using guided resonant modes with Bragg mirrors outside of the PC, allows for multiple linear arrays with 1D radiative behavior whose light can be extracted through one
of the mirrors. This setup could be used to implement in a massive parallel way many protocols designed for 1D arrays of atoms.

\indent {\it Acknowledgements.--} We acknowledge fruitful discussions with A. Gonz\'alez-Tudela and J.I. Cirac.
This work has been supported by the EU 
through the H2020 Project QuProCS (Grant Agreement 641277), by MINECO/AEI/FEDER through projects NoMaQ FIS2014-60343-P, QuStruct FIS2015-66860-P
and EPheQuCS FIS2016-78010-P. FG acknowledges funding from `Vicerectorat d'Investigaci\'o  i Postgrau'  of the UIB.

\end{document}